# Advanced Technology in Speech Disorder Therapy of Romanian Language

Mirela Danubianu, Iolanda Tobolcea, and Stefan Gheorghe Pentiuc

**Abstract**— One of the key challenges of the society development is related to public health and one of its specific targets includes better treatments of diseases. It is true that there are affections which by their nature do not endanger the life of a person, but they may have negative implications during his/her lifetime. Various language or speech disorders are part of this category. Discovered and treated in time, they can be corrected, most often in childhood. Because the Romanian language is a phonetic one that has its own special linguistic particularities, there is a real need to develop advanced information systems, which can be used to assist and help specialists in different speech disorders therapy. The aim of this paper is to present a few CBTS developed for the treatment of various language and speech disorders specific to the Romanian language.

—————————— ◆ ——————————

## 1 INTRODUCTION

SPEECH and language disorders are diseases that do not endanger a person's life, but can cause many troubles in his/her interpersonal relations. Most the speech disorders are met during childhood and many of them are solved spontaneously, once with the individual's development. There are nevertheless disorders that require the specialist's intervention in order to be corrected or at least improved.

Since the researches carried out revealed the fact that an important number of persons suffer from such disorders, in the past few years, efforts have been made in order to find some efficient therapy programs, resorting to informatics technology.

Various roles have been assigned to the informatics systems: from the simple replacement of drawn didactic materials with images, sometimes 3D models, presented on the computer monitor, to considering the computer a real assistant for the speech therapist, helping thus in evaluating the situation of an individual and in establishing the diagnosis, in projecting a personalized treatment plan and in the treatment follow-up.

The aim of this paper is to present a few CBTS developed for the treatment of various language and speech disorders specific to the Romanian language.

Section 2 refers to the characteristics of the language disorders and to their possible implications. Section 3 approaches some deficient aspects related to the use of ICT in the speech therapy practice and Section 4 presents some projects of this kind.

A short presentation of the Echophone, Logoped 1.0, and Terapers systems is carried out in Section 5.

Also, as an answer to the challenge aiming to the analysis of the data collected in the Terapers program by data mining techniques, this section presents the Logo-DM program, which is currently in the implementation stage.

## 2 CHARACTERISTICS AND EDUCATIONAL IMPLICATIONS OF SPEECH DISORDER

Speech and language impairments address problems in communication and related areas such as oral motor function [1]. These disorders range from simple sound substitutions to the inability to understand or use language or use the oral-motor mechanism for functional speech and feeding. Some causes of these problems include neurological disorders, mental retardation, hearing loss, brain injury, drug abuse, physical impairments such as cleft lip or palate, and vocal abuse or misuse. Speech disorders refer to difficulties in producing speech sounds or problems related to the voice quality and they might be characterized by an interruption in the flow or rhythm of speech, such as stuttering, which is called dysfluency or by the problems with the way sounds are formed, called articulation or phonological disorders. Also there might be difficulties related to the volume or quality of the voice. Sometimes there may be a combination of several problems. People with speech disorders have troubles using some speech sounds, which can also be a symptom of a delay. They may say "lac" when they mean "rac" or they may have trouble using other sounds like "c" or "g." Listeners may have trouble understanding what someone with a speech disorder is trying to say.

All communication disorders have the potential to isolate individuals from their social and educational environment. While many speech and language patterns can be called "baby talk" and are part of a young child's normal development, they can become problems if they are not outgrown as expected. In this way an initial delay in speech and language or an initial speech pattern can become a disorder likely to cause learning difficulties. Because of the way the brain develops, it is easier to learn language and communication skills before the age of 5. When children have muscular disorders, hearing problems or developmental delays, their acquisition of speech, language and related skills is often affected. Consequently, it is essential

————————————
- M. Danubianu is with the "Stefan cel Mare" University of Suceava, Faculty of Electrical Engineering and Computer Science
- I. Tobolcea is with the "Alexandru Ioan Cuza" University of Iasi, Faculty of Psichology and Education Science
- St. Gh. Pentiuc is with the "Stefan cel Mare" University of Suceava, Faculty of Electrical Engineering and Computer Science

to find appropriate timely intervention.

The logopaedic intervention proposes some specific objectives such as detection, complex assessment and identification of language and speech disorders of preschoolers and small scholars' children and targeting logopaedic therapy to correction, recovery, compensation, adaptation and social integration. This latter goal involves the application of a personalized therapy to each child or group of children with similar characteristics, therapy adjusted according their disease severity and directed towards eliminating the causes that generated the speech impairment.

## 3 THE USE OF INFORMATION AND COMMUNICATION TECHNOLOGY IN SPEECH THERAPY PRACTICES

Information and Communication Technology can help children whose physical conditions make communication difficult. For example, the use of electronic communication systems allows nonspeaking people and people with severe physical disabilities to engage in the give and take of shared thought.

ICT can be used in speech therapy as a real clinical tool. The speech therapist can use harmoniously his/her clinical competences and the technical means in order to help their patients, orienting the evaluation, the treatment and the constant feedback in a more flexible and modern way.

The computer used in speech therapy can help in diagnosing the speech disorders, can produce audio-visual feedback during the treatment and monitor and evaluate the therapeutic progress. Also, it provides some sets of practical exercises for the patients who are not under the direct supervision of a speech therapist.

The use of Computer Based Speech Therapy (CBST) systems in various stages of speech therapy determines a new psychological and pedagogic situation, by creating a special learning environment, by facilitating a new, superior method for information recovery. The enrichment of the material resources as support for an efficient and rapid learning by means of new tools of communication such as computers, cultivate the children's interest in the therapeutic activity. By systematically presenting and giving new, rich and well selected information that can be reproduced in their natural environment and dynamics, the computer supports the children's curiosity for ongoing knowledge acquisition and increases their motivation for learning. Therefore, the computer constitutes not only a new method of communication, but it also facilitates affective states, it transmits emotions, feelings, attitudes, contributing to the improvement of the children affective life.

Consequently, some of the advantages of using the CBST are:
1. the possibility to detect aspects impossible or extremely difficult to realize by other means and to separate or recompose phenomena that are imperceptible by other means;
2. the capacity to playback the content extremely accurately and to allow immediate playback of the information or as often as necessary;
3. they are attractive for children due to the original aspects involved, as well as to the aesthetic way of presenting the information;
4. they always maintain the child in the present day situation.

To conclude, among the arguments supporting the use of ICT in increasing the efficiency of speech therapy, we can mention:
1. the recordings of the verbal material provide that "language immersion" necessary for acquiring a correct pronunciation;
2. they constitute tools of self-control of the errors made and the progress achieved, putting the child in the situation to hear himself / herself and to judge from the outside;
3. they provide the speech therapist with tools for controlling and evaluating the efficiency of the proposed didactic strategy, helping to reconsider the therapeutic demarche;
4. they constitute an important aid for the therapist in the analysis of the speech therapy sessions, the responses received from the persons with speech disorders, managing to appreciate the efficiency of the methods, the means used in reaching the established goal;
5. they broaden the knowledge horizon of the children, develop logical thinking, and develop affectivity and speech;
6. they increase the efficiency of the therapy, create a pleasant, relaxed, and attractive climate, determining a better state of mind; the desire for success interferes, self trust and the trust in one's own corrective capacities.

In fact a CBST constitutes the most complex method that comprises the audio-visual techniques. It has the great advantage to realize some educational and instructive software programs; it helps and it increases the efficiency of the didactic activity.

## 4 INTERNATIONAL PROJECTS REGARDING THERAPEUTIC SOFTWARE FOR SPEECH DISORDERS

Lately, therapeutic software used for correcting various speech disorders has been elaborated. Some software programs are simple and produce a single type of visual and auditory feedback, while others are extremely complex, allowing a sustained training, realized for several aspects of speech. The integration of audio-visual tools in speech therapy together with the therapist activity, contribute to a faster achievement of the therapeutic goals, such as correct pronunciation, speech development in general, improvement of the affective and emotional life, stimulation of imagination.

Here we present such projects, achieved on the international level. The priorities are represented by developing information systems that will allow the elaboration of personalized therapeutically paths. The following main directions are considered: development of expert systems personalizing the therapeutic guides to the child's evolution and the evaluation of the motivation and progresses that the child's achieves.

The OLP (Ortho-Logo-Paedia) project [2] for speech therapy started in 2002; the project is financed by the EU and it is a complex project, involving the Institute for Language and Speech Processing in Athens and seven other partners from the universitary and medical domains. It aims to accomplish a three – module system (OPTACIA, GRIFOS and TELEMACHOS) capable of interactively instructing the children suffering from dysarthria (difficulty in articulating words due to disease of the central nervous system). The proposed interactive environment is a visual one and is adapted to the subjects' age (games, animations). The audio and video interface with the human subject will be the OPTACIA module, the GRIFOS module will make pronunciation recognition and the computer aided instruction will be integrated in the third module – TELEMACHOS.

An interesting project is STAR – Speech Training, Assessment, and Remediation [3], started in 2002, a project which is still in the development phase. The members (AI. duPont Hospital for Children and The University of Delaware) aim to build a system that would initially recognize phonemes and then sentences. This research group offers a voice generation system (ModelTalker) and other open source applications for audio processing.

Speechviewer III developed by IBM [4] creates an interactive visual model of speech while users practice several speech aspects (e.g. the sound voice or special aspects from current speech).

The ICATIANI device developed by TLATOA Speech Processing Group, CENTIA Universidad de las Américas, Puebla Cholula, Pue. México uses sounds and graphics in order to ensure the practice of Spanish Mexican pronunciation [5]. Each lesson explains sounds pronunciation using the facial expression with a particular accent on specifying articulation points and the position of the lips. The system includes several animated faces, each of them showing the correct method of vocal pronunciation and providing feedback to the child answers. In this case, if the child's pronunciation matches the system one, the child is rewarded by a smile or warned by a sad face otherwise.

As we noted, at international level, the software applications are quite expensive (500$-1500$) and inappropriate for the specific phonetics of the Romanian language.

## 5 ICT IN ROMANIAN SPEECH THERAPY PRACTICES

Taking into consideration the fact that Romanian language is a phonetic one that has its own special linguistic particularities, we consider that there is a real need for the development of advanced technology systems which can be used in the therapy of different speech or language problems.

At the national level, some researches have been conducted on the therapy of speech impairments, most of which are focused on traditional areas such as voice recognition, voice synthesis and voice authentication. The studies carried out at the Faculty of Psychology and Education Science from "Al. I. Cuza" University of Iasi and at the Faculty of Electrical Engineering and Computer Science from "Stefan cel Mare" University of Suceava have lead to developing systems which assist the speech therapists in their efforts to design a better therapy or which provide feedback regarding the oral fluency.

### 5.1 Echophone

Echophone is a system used for achieving the delayed auditory feedback (DAF) [5]. Its construction started from the hypothese, verified during the research in therapy, that by listening to their own speech, the logoneurotic patients can correct the rhythm of their speech, which influences positively the speech fluency.

The Ecophone allows:
1. a clear and correct reception of the sounds emitted by the subject, that creates a real image of the subject about his/her pronunciation in comparison with the same sounds pronounced by the speech therapist;
2. to replace the fast rhythm of speech, with a normal speech by the effectuation of some speech therapy exercises adequate to the structure of the logoneurotic patient and to the evolution of the treatment;
3. the increase in the will and the energy of patients to correct as rapidly as possible the speech disorders by realizing the progresses made in the therapy

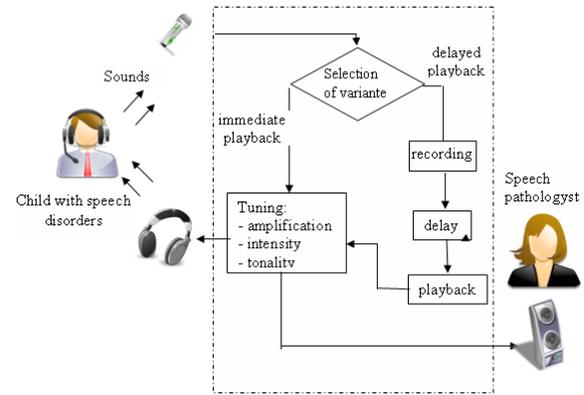

Fig. 1. The Echophone architecture

Fig. 1 presents the architecture of the Echophone. In front of the microphone, the child utters the words rhythmically, rarely, in order to hear them like an echo. The child has the possibility to realize his/her own articulation, and control the speech and the voice. Spontaneously, during the apparition of the verbal dysfluences or spasms, after a certain break, the logoneurotic patient recommences to speak pronouncing the words fluently, correctly.

We can conclude that by using the Echophone, great performances can be achieved in re-educating the speech of the logoneurotic patients by delaying the auditory feedback. Also, the echophone contributes to realizing the progresses made by the logoneurotic patient in speech, increasing the courage and trust in the capacity to express himself/herself correctly and fluently.

### 5.2 Logoped 1.0

It is created for the treatment of logoneurosis, but it is also useful for the therapeutic activities in the treatment of dyslexic-dysgraphic disorders, providing a vast lexical material grouped on several sections: reading the syllables and the words, reading sentences, reading phrases and reading texts. The colorful presentation of words and sentences, their display on the monitor, the possibility to select the degree of difficulty of the exercises make the reading activity supervised by the speech therapist much more attractive for the pupil [5].

This program has the following advantages, compared to the classical, traditional activity:
1. the material comprised is presented in a gradation meant to fulfill all the requirements of a progressive treatment;
2. it arouses much more the interest and the dynamism for participating in the therapeutic activity, considering that the writing is colored, the reading rhythm is followed on the computer monitor, the results are displayed on the right side of the screen, stimulating the will of the subject in outgrowing the present performances;
3. the performances established by the calculator are reached according to the physical and psychical capacity of each subject;
4. an active relation „subject-computer" is established by the appreciations given by the computer for the performances reached by the subject;
5. the promise of a reward, stimulation provided by the computer for the subject's performances [5].

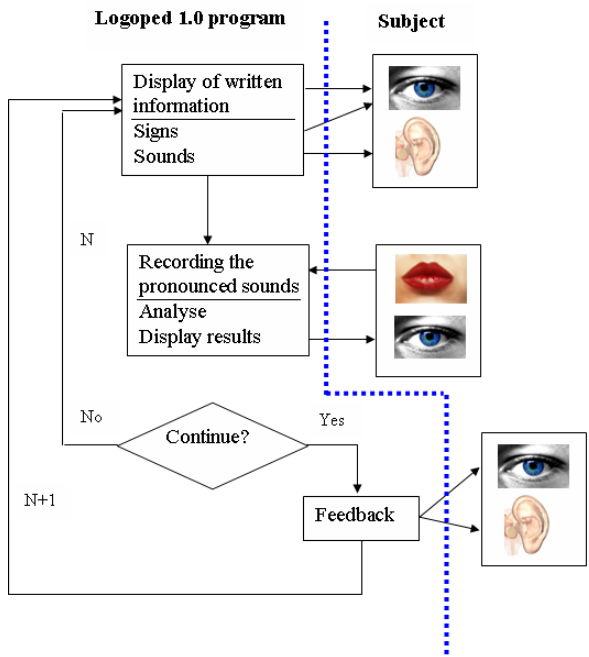

Fig. 2. The Logoped 1.0 functional schema

## 5.3 Terapers

The information systems with real time feedback that address pathological speech impairments are relatively recent due firstly to the amount of processing power they require [6]. The progress in computer science allows at the moment for the development of such a system with low risk factors. Children pronunciation is also used to enrich the existing audio database and to improve the current diagnosis system's performances.

The personalized therapy system of dyslalia - Tearpers has reached some specific objectives [7] [8]:
1. initial and ongoing therapy evaluation of children and identification of a modality for standardizing their progresses and regresses (at the level of the physiological and behavioral parameters);
2. rigorous formalization of an assessing methodology and development of a pertinent database in this area;
3. the development of an expert system for the personalized therapy of speech impairments that allows designing a training path for pronunciation, individualized according to the speech disorder category, previous experience and the previous evolution of the child's therapy;
4. the development of a therapeutically guide that allows mixing classical methods with the adjuvant procedures of the audio-visual system

5. the design and the achievement of a database that contains the child's dates, the set of exercises and the results obtained by the child.

The high degree of complexity of the project results from the high number of different research areas involved: artificial intelligence (learning expert systems, pattern recognition), virtual reality, digital signal processing, digital electronic (VLSI), computer architecture (System on Chip, embedded device) and psychology (evaluation procedures, therapeutic guide, experimental design for validation).

The architecture of TERAPERS is presented in Fig. 3. The system contains two main components: an intelligent system installed on each office computer of the speech therapists and a mobile system used as a friend for the child therapy. The two systems are connected [9].

The intelligent system is the fix component of the system and it is installed on each computer in the speech therapist's office.

This system includes the following parts:
1. a children information management module
2. an expert system that will produce inferences based on the data presented by the evaluation module.
3. a virtual module of the mouth, that would allow the presentation of every hidden movement occurring during speech
4. an exercises management module that allows for the creation or modification of the exercises corresponding of various stages of therapy and grouping them in complex issues

The mobile device of personalized therapy has two main objectives. It is used by the child in order to resolve the homework prescribed by the speech therapist and delivers to the intelligent system a personalized activity report of the child.

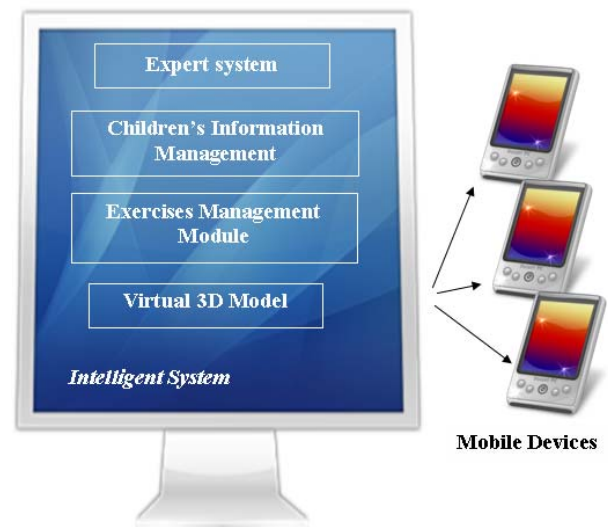

Fig. 3. TERAPERS Architecture

The main stream activities of intelligent system from TERAPERS are presented in Fig. 4. All these activities are materialized in a consistent volume of data stored in a relational database [10].

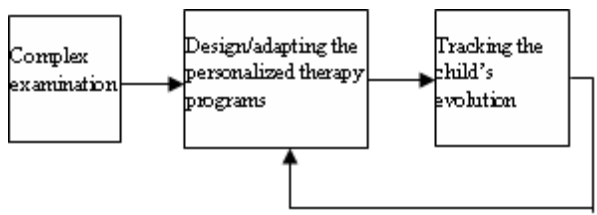

Fig. 4. The functional schema of TERAPERS

Adapting the therapy programs involves a complex examination of children and recording the relevant data related to personal and family anamnesis. Complex examination of how the children articulate the phonemes in various constructions allows for a diagnosis and classification in a given category of severity. Anamnesis data collected may provide information about various causes that may negatively influence the normal development of language. It contains historical data and data provided by the cognitive and personality examination.

To design an appropriate schema for personalized therapy, we provide to the system data such as number of sessions/week, exercises for each phase of therapy and the changes of the original program according to the patient evolution. In addition, the report downloaded from the mobile device collects data on the efforts of child self-employment. These data refer to the exercises done, the number of repetitious for each of these exercises and the results obtained.

The tracking of child's progress materializes the data indicating the moment of assessing the child and his status at that time.

Starting March 2008 the system is currently used by the therapists from Regional Speech Therapy Center of Suceava and it has already demonstrated its efficiency.

### 5.4 Logo-DM

The development and use of information technology in order to assist and follow speech disorder therapy allowed the researchers to collect a considerable volume of data. Increased volume of data available did not lead immediately to a similar volume of information to support the decisions of effective therapy, because the classical methods of data processing are not applicable in such cases.

We think these data can be the foundation of data mining processes that show interesting information for the design and adaptation of different therapies in order to obtain the best results in conditions of maximum efficiency.

Data mining involves the analysis application on large volumes of data using algorithms which produce a particular enumeration of patterns from such data. Results obtained through the application of appropriate methods of data mining can provide answers to two broad categories of problems: prediction and description [11].

The idea of trying to improve the quality of logopaedic therapy by applying some data mining techniques started from TERAPERS project developed within the Center for Computer Research in the University "Stefan cel Mare" of Suceava [9]. Data collected in this system together with data from other sources (eg demographic data, medical or psychological research) may be the set of raw data that will be the subject of data mining. To this end, we have proposed the development of Logo-DM system.

In order to obtain useful patterns from these data, we decided to use the following methods:

- classification, to place the people with different speech impairments in predefined classes. We use a classification based on the information contained in many predictor variables, such as personal or familial anamnesis data or data related to lifestyle, to join the patients with different segments.
- clustering, to group people with speech disorders on the basis of similarity of different features. It is not based on the previous definition of groups but it helps the therapists to understand who their patients are. Clustering aims to find subsets of a predetermined segment, with homogeneous behavior towards various methods of therapy that can be effectively targeted by a specific therapy.
- association rules, to find out associations between different data which seem to have no semantic dependence. An important task of the association is to determine why a specific therapy program has been successful on a segment of patients with speech disorders and on the other it was ineffective.

A first version of the proposed architecture for this system is presented in Fig. 5.

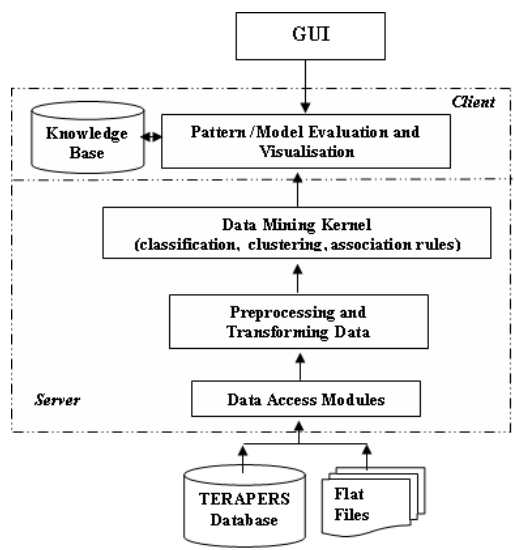

Fig.5. Logo-DM architecture

On the client side there is the user interface (GUI) which allows the user to communicate with the system in order to select the task to perform, to select and submit the datasets on which data mining needs to be applied. Pattern evaluation and pattern visualization are performed also on the client. The knowledge base is the module where the background knowledge is stored.

The more difficult computational tasks of data mining operations are carried out on the server. Here, the data mining kernel contains modules able to perform classifications, clustering and association rule detection. Supplementary the preprocessing data module allows data to become suitable for applying data mining algorithms.

## 6 CONCLUSION

The elaboration and implementation of the computer based speech therapy systems allow for the increase in the efficiency of speech disorders treatment by equipping the speech therapy medical offices with therapeutic software and computers which are able to apply supplementary informatics means.

By using a mobile device, as it is possible with TERAPERS, the child will be able to continue the therapeutic exercises at home, without the speech therapist. This can eliminate the therapeutic losses caused by pauses in the supervised practice during the holidays and indirectly, this could lead to a financial benefit for the families with dyslalic children.

It should be mentioned that when using that kind of systems in some stages of the pre-school speech therapy, we should take into consideration aspects such as: the age peculiarities of the person with speech disorders and the speech peculiarities, the personality of the person with speech disorders, the level of mental and of language development and the capacity to receive and interpret audio and visual stimuli. Ignoring these demands can lead to shortcomings and limits in the use CBST like apathy, passivity, psychic fatigue and behavior standardization.

**Mirela Danubianu** received a MS in Computer Science at University of Craiova, Faculty of Electrical Engineering (1985 – Automatizations and Computers) and other in Economics at Stefan cel Mare University of Suceava, Faculty of Economics (2009 - Management). She is PhD in Computers Science at "Stefan cel Mare" University of Suceava, Faculty of Electrical Engineering and Computers Science (2006 - Contributions to the development of data mining and knowledge methods and techniques).. Now she is lecturer at "Stefan cel Mare" University of Suceava. Faculty of Electrical Engineering and Computers Science, Her current research interests include databases theory and implementation, data mining and data warehousing, application of advanced information technology in economics and health care. She has (co)authored 7 books and more than 25 papers, more than 15 conference participations, member of the International Program Committee of 3 conferences.

**Iolanda Tobolcea** is Associate Professor at the Faculty of Psychology and Educational Sciences, "Alexandru Ioan Cuza" Univesity of Iasi, Department of Clinical Psychology and Special Education. She is PhD in Psychology (1997- The Use of Modern Audio-Video Techniques in the Therapy of stuttering in Children of School Age).Current and past works: speech and language disorders therapy, psychotherapy, pedagogical counseling, psycho impaired intellect, experimental psychology. She has done studies and research activity on the methods, processes and techniques used for the recovery of persons with disabilities (language, sensory, motor, intellectual etc.). She has done many studies on the therapy of language disorders, the results of which were gathered in 6 books and more than 30 papers and conference participations, focused mainly on the development of different software for the therapy of rhythm and fluency disorders in speech, writing, reading or articulation of sounds.

**Stefan Gheorghe Pentiuc** is professor at "Stefan cel Mare" University of Suceava. He received his Ph. D. degree in 1993 in Computer Science and Engineering at the "Politehnica" University of Bucharest based on a thesis on pattern recognition. Current and past research interests: pattern recognition, distributed artificial intelligence, human computer interaction. He is the head of the Research Center in Computer Science  and the dean of Faculty of Electrical Engineering and Computer Science from "Stefan cel Mare" University of Suceava.